\documentclass[sn-mathphys-num]{sn-jnl}% Math and Physical Sciences Numbered Reference Style 

\usepackage{graphicx}%
\usepackage{multirow}%
\usepackage{amsmath,amssymb,amsfonts}%
\usepackage{amsthm}%
\usepackage{mathrsfs}%
\usepackage[title]{appendix}%
\usepackage{xcolor}%
\usepackage{textcomp}%
\usepackage{manyfoot}%
\usepackage{booktabs}%
\usepackage{algorithm}%
\usepackage{algorithmicx}%
\usepackage{algpseudocode}%
\usepackage{listings}%
\begin{document}

\title[]{Solar peculiar motion inferred from dipole anisotropy in redshift distribution of quasars appears to lie along the Galactic Centre direction}

\author*{\fnm{Ashok K.} \sur{Singal}}\email{ashokkumar.singal@gmail.com}

\affil{\orgdiv{Astronomy and Astrophysics Division}, \orgname{Physical Research Laboratory}, \orgaddress{\street{Navrangpura}, \city{Ahmedabad}, \postcode{380009}, \state{Gujarat}, \country{India}}}
\abstract{According to the Cosmological Principle an observer stationary with respect to the comoving coordinates of the expanding universe 
should find the redshift distribution of distant quasars to be isotropic. However, the observed redshift distribution in a large sample of 1.3 million quasars shows a significant dipole anisotropy.
A peculiar motion of the observer could introduce such a dipole anisotropy in the observed redshift distribution. However, the motion inferred therefrom turns out to be not only many times the peculiar motion estimated from the anisotropy in the Cosmic Microwave Background (CMB), but also nearly in a direction at a right angle. The Solar peculiar motion, in fact, turns out to be, quite unexpectedly, in the direction of the Galactic Centre. Such a statistically significant discrepancy in peculiar motion, derived by different methodologies, could imply a violation of the cosmological principle, a cornerstone in the foundation of the standard model.}


%%\pacs[JEL Classification]{D8, H51}

%%\pacs[MSC Classification]{35A01, 65L10, 65L12, 65L20, 65L70}

\maketitle

%%%%%%%%%%%%%%%%%%%%%%%%%%%%%%%%%%%%%%%%%%%%%%%%%
\section{Introduction}\label{S1}
A dipole anisotropy in the Cosmic Microwave Background (CMB) in the direction RA$=168^{\circ}$, Dec$=-7^{\circ}$ (in galactic coordinates, $ l=264^{\circ}, b=48^{\circ}$) seen by COBE, WMAP and Planck satellites has been interpreted as due to a peculiar velocity 370 km s$^{-1}$ of the solar system in that direction \cite{1,2,3}.  
%(Lineweaver et al. 1996; Hinshaw et al. 2009; Akrami et al. 2018) 
Such a peculiar motion should also give rise to similar dipoles in the sky distribution of distant radio sources and some initial searches for the radio source dipoles, because of the large statistical uncertainties, concluded that the results were more or less in agreement with the CMB dipole \cite{20,21}. 
%(Baleisis et al. 1998; Blake \& Wall 2002).
However, from dipole anisotropies in the sky brightness arising from discrete sources as well as their number counts in the sky, it was shown \cite{4} that  
the solar peculiar motion is $\sim 1600$ km s$^{-1}$, a factor of about 4 higher than the CMB value at a statistically significant level, though the direction might within errors be in agreement with the CMB dipole. 
%(Singal 2011; Rubart \& Schwarz 2013; Tiwari et al. 2015; Colin et al. 2017; Bengaly et al. 2018; Singal 2019a,b; Secrest et al. 2021; Siewert et al. 2021; Singal 2021a,b) 
Subsequently many independent groups employing distant radio, infrared or optical sources corroborated the claim that the solar peculiar motion may be much larger than that inferred from the CMB, though along  the same direction \cite{5,6,7,8,9,11,Si21,Sie21,Se22,Wa23,Si24a,Si24b,Pa24}. 
%(Gibelyou \& Huterer 2012; Rubart \& Schwarz 2013; Tiwari et al. 2014; Colin et al. 2017; Bengaly, Maartens \& Santos 2018) 

Here we strive to determine in an independent and rather more direct method, our peculiar motion from a dipole anisotropy in the redshift distribution on the sky of a large sample of quasars.
A distinct advantage with the redshift dipole is that it can provide a direct estimate of the solar peculiar velocity with respect to the reference frame of the 
quasars. It also obviates the need for looking for local clustering or structure formations on large scales that might otherwise masquerade, for instance, as a number count dipole asymmetry purportedly due to observer's motion in other indirect methods (see e.g. \cite{29,34a,19,Oa24}). 
%Rubart, Bacon \& Schwarz 2014; Tiwari \& Nusser 2016; Rameez et al. 2018).
%\section{The quasar Sample}\label{S2}
A preliminary investigation of the dipole anisotropy in the redshift distribution of $\sim 10^5$ quasars  
%a subset of DR12Q  catalogue \cite{27},
%(P\^aris et al. 1917), 
indicated the peculiar velocity to be $\sim 6$ times the CMB value but in an opposite direction \cite{10}. The sample, however, was not only small and it covered the sky also in a small number of patches. But now, an order of magnitude larger, all-sky spectroscopic quasar catalogue,   
%The sample we employ for our study is 
Quaia, is publicly available  \cite{St24}. The catalogue is drawn originally from the 6,649,162 quasar candidates identified by the Gaia mission \cite{Ga23b} that have redshift estimates from the space observatory's low-resolution blue photometer/red photometer slitless spectrograph \cite{De23}. The above Gaia candidates are combined with unWISE infrared data (based
on the Wide-field Infrared Survey Explorer survey \cite{18}) to construct a quasar catalogue useful for cosmological and astrophysical
studies. Cuts were applied based on proper motions and colors, reducing the number of contaminants by
approximately four times and the redshifts were improved by training a k-Nearest Neighbour model on SDSS redshifts \cite{Ly20}.
%and achieve estimates on the G < 20.0 sample with only 6% (10%) catastrophic errors with |Δz/(1 + z)| > 0.2(0.1), 
%a reduction of approximately three times (approximately two times) compared to the Gaia redshifts. 
The final catalogue, containing 1,295,502 sources to a Gaia G-band magnitude limits $m_{\rm G}< 20.5$, is presently the largest-volume spectroscopic quasar catalogue 
%catalog with an additional magnitude cut,
%QUAIA, the Gaia-unWISE Quasar Catalog, is a 
and presumably, may be a highly competitive sample for cosmological large-scale structure analyses \cite{St24}. Although a Bayesian analysis of the Quaia sample found a dipole that seems to be in agreement with the CMB dipole \cite{Mi24}, however, a direct computation of the dipole from asymmetries seen in the source number counts gave a dipole 3-4 times as large as the CMB dipole though in the same direction \cite{Si24b}. Here we employ the Quaia sample and search for a dipole anisotropy, if any, in the angular distribution of the redshifts of these quasars in the sky, in order to determine Solar peculiar motion.
 %(Storey-Fisher et al. 2023). 
%--------------------------------------------
%\section{QUAIA sample of quasars}
%QUAIA, the Gaia-unWISE Quasar Catalog, is a publicly available, an all-sky spectroscopic quasar sample, which may be a highly competitive sample for cosmological large-scale structure analyses (Storey-Fisher et al. 2023). The sample is drawn from the 6,649,162 quasar candidates, identified by the Gaia mission (Gaia Collaboration 2016), released in Gaia DR3 (Gaia Collaboration 2023a,b). Further, all Gaia
%quasars have been cross-matched with those from the Wide-field Infrared Survey Explorer (WISE; Wright et al. 2010), to also provide photometric information in the W1 and W2 infrared bands. 
%The QUAIA catalogue is available in two versions with different Gaia G-band magnitude limits, the full  $m_{\rm G}< 20.5$ version containing 1,295,502 quasars, and a reduced but cleaner version with 755,850 quasars, which is just a subset of the larger catalogue with an additional magnitude cut, $m_{\rm G}< 20.0$. 

%---------------------------------------------------
\section{Method}\label{S2}
By the  cosmological principle (CP) distant quasars should have isotropic distribution of cosmological redshifts for an observer at rest with respect to comoving coordinates of the expanding universe.
However, due to a peculiar velocity $v$ (assuming a non-relativistic motion with $v\ll c$ \cite{1,2,3}) of the Solar system, an observer on Earth will find a quasar lying at an angle $\theta$ with respect to the direction of motion, as seen in observer's frame, to have a redshift
\begin{equation}
\label{eq:1}
(1+z)= (1+z_{\rm o})/\delta=(1+z_{\rm o})[1-(v/c)\cos\theta] 
\end{equation}
where $z$ is the redshift of the quasar measured by the observer, $z_{\rm o}$ is the cosmological redshift of the quasar, and $\delta=[1-(v/c)\cos\theta]^{-1}$ is the Doppler factor due to the observer's  non-relativistic motion. 

The observed redshift will then display a dipole pattern
\begin{equation}
\label{eq:2}
 z= z_{\rm o}+{\cal D}\cos\theta,
\end{equation}
with the dipole component ${\cal D}=-(1+z_{\rm o})(v/c)$, arising from the peculiar velocity $v$  of the observer. From the CP, the isotropic distribution of $z_{\rm o}$ implies an average of the observed redshift $z$ over different directions will cancel the dipole component, i.e., $\bar{z}={z}_{\rm o}$. 
The redshift distribution of distant quasars may thus directly yield an estimate of the Solar peculiar motion. 

The procedure to be followed is straightforward. We first divide the sky into a grid of, say, $10^\circ \times 10^\circ$ bin size with $422$ pixels and with respect to the centre of each of the, say $i^{th}$ pixel, we calculate the polar angle $\theta$ for each of 1.3 million quasar positions in the sky.
Then a plot of the observed redshifts, $z$, against $x=\cos\theta$ for these quasars should yield a linear correlation, $z=a_{\rm i}+b_{\rm i}x$. Therefore, making a least square fit of a straight line to the ($x,z$) data on quasars in our sample we can get ${z}_{\rm o}$ and ${\cal D}$ (with $z_{\rm o}=a_{\rm i}$ and ${\cal D}=b_{\rm i}$), which yields the magnitude of the solar peculiar velocity $v_{\rm i}$ as
\begin{equation}
\label{eq:3}
\frac{v_{\rm i}}{c}=\frac{-{\cal D}}{1+z_{\rm o}}=\frac{-b_{\rm i}}{1+a_{\rm i}}.
\end{equation}
This, however, yields only a projection of the peculiar velocity $v_{\rm i}$ in the direction of $i^{th}$ pixel, which should have a $\cos\psi_{\rm i}$ dependence where $\psi_{\rm i}$ is a polar angle of $i^{th}$ pixel with respect to the actual direction of the Solar peculiar motion. A contour plot of the 422 values should yield both magnitude and direction of the peculiar motion.
We also make a 3-d $\cos\psi$ fit for each of the $n=422$ positions for the remaining $n-1$ $v_{\rm i}$ values, and compute the chi-square value for each of these $n$ fits. 
The location where a minimum of these $n$ chi-square fits occurs should yield the optimum value of the peculiar velocity direction, which should lie close to the maximum in the contour plot. 
For convenience of a comparison with the CMB dipole, we express the peculiar velocity  in units of the CMB value as $v=p\times 370$ km s$^{-1}$, where $p=1$ means velocity equal to the CMB value of 370 km s$^{-1}$ while $p=0$ will imply a nil peculiar motion. 

In order to verify the procedure and validate our computer routines we performed a large number of Monte Carlo simulations. In each trial, while keeping the original quasar positions, so as to keep the sky coverage unaltered, we first redistributed the observed redshifts among quasars randomly and then on each of these randomly obtained quasar redshift we superposed the effects of an assumed realistic peculiar velocity vector. On this mock catalogue we applied our procedure to recover the peculiar velocity vector and compared with the value assumed in that particular simulation. This way we not only verified our technique but this also allowed us from one thousand such simulations to estimate errors in the derived peculiar velocity, both in magnitude and direction.
We also tried finer grids of $2^\circ \times 2^\circ$ size bins with $10360$ cells which, however, made barely perceptible difference in our results. 
%---------------------------------------------------
\begin{figure*}
\includegraphics[width=\linewidth]{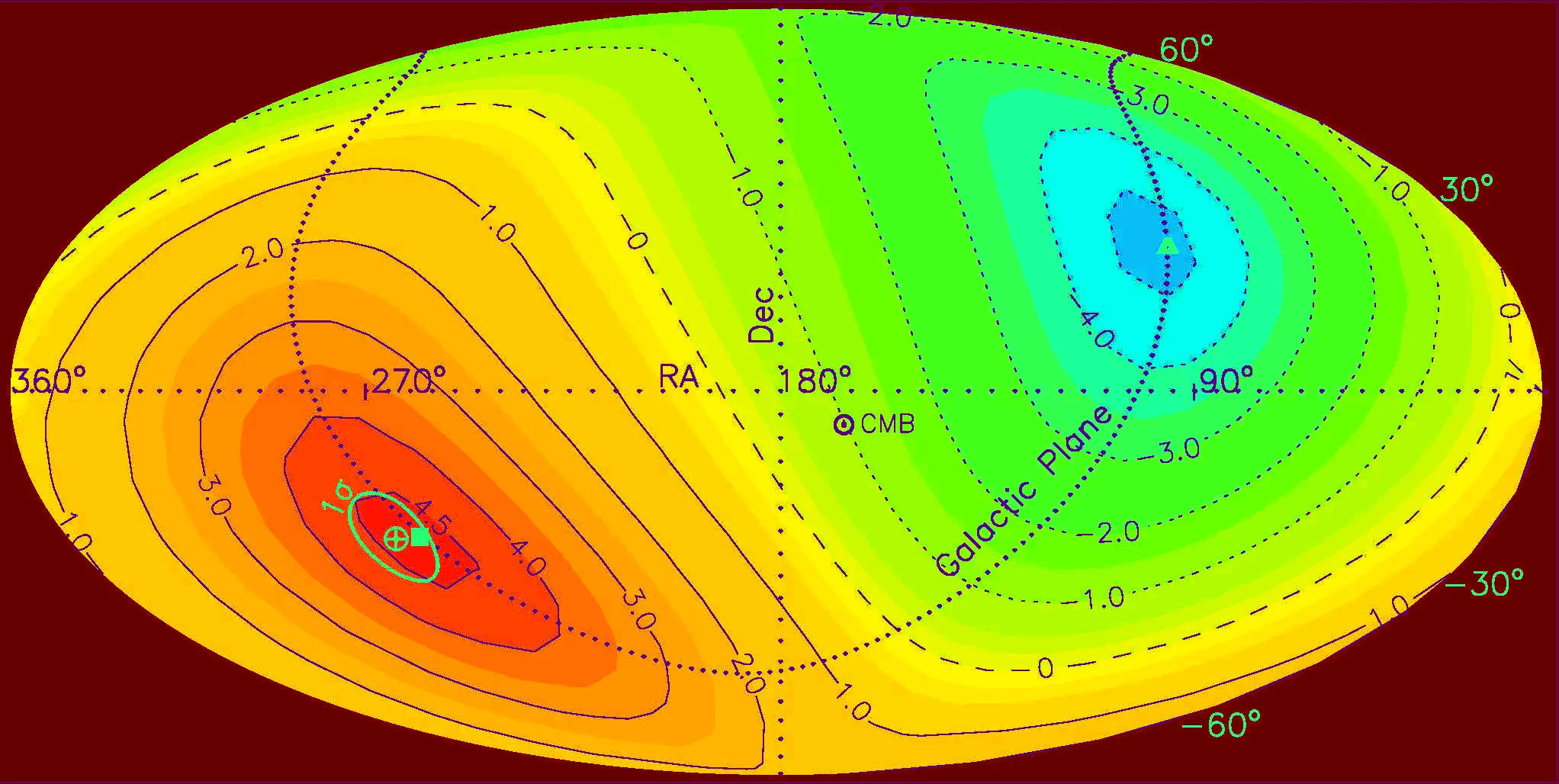}
\caption{A Hammer--Aitoff equal-area projection coloured map in equatorial coordinates, of the peculiar velocity components, estimated for 422 directions ($10^\circ\times10^\circ$ grid points) in the sky from the Quaia $m_{\rm G}<20.5$ data. Overlaid are some representative contours, depicting amplitudes of the peculiar velocity, in units of the CMB value of 370 km s$^{-1}$. The~horizontal and vertical axes denote RA and Dec. Positive component values are shown by continuous contour lines; the dashed curve represents zero amplitude of the peculiar velocity component while the negative component values are shown by dotted lines. 
The~optimum pole direction is expected to be in the vicinity of the highest contour value (in dark red colour) and is represented by the symbol $\oplus$ (in light blue), 
%lying at the position where a 3-d $\cos\psi$ fit yields a minimum chi-square, 
while $1\sigma$ errors are indicated by a light blue-coloured ellipse around it.
Symbol $\odot$ indicates the CMB pole position. Also shown is the Galactic Plane (thick-dotted curve) with the symbol $\blacksquare$  (in light blue) indicating position of the Galactic Centre while $\blacktriangle$ (in light blue) indicates  position of the Galactic Anticentre.\label{F1}}
\end{figure*}
%%---------------------------------------------------
\begin{figure}
\begin{center}
\includegraphics[width=8cm]{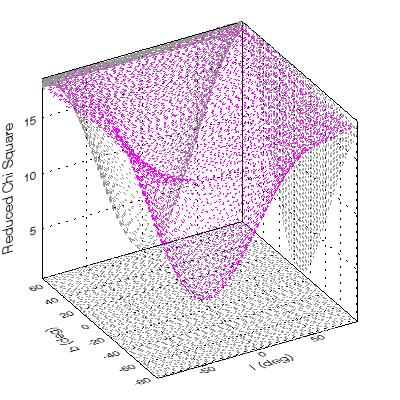}
\caption{A 3-D plot of the reduced chi-square ($\chi^2_\nu$) values (in violet colour), from the cosfit routine for various pixels in the sky for the whole Quaia data  ($m_{\rm G}<20.5$). The horizontal plane shows the direction in sky in galactic coordinates ($l,b$). The position of the minimum of the reduced chi-square is determined at $l=3^{\circ}$ and $b=-6^{\circ}$ from the 2-D projections, shown in light grey.}
 \label{F1a}
\end{center}
\end{figure}
%---------------------------------------------------

\begin{figure*}
\includegraphics[width=\linewidth]{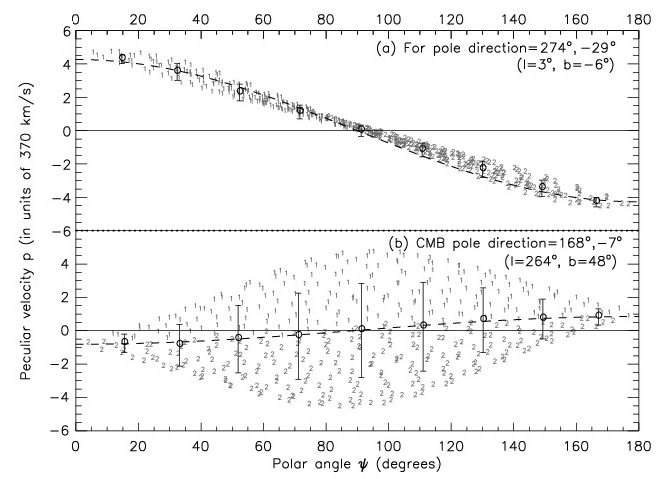}
\caption{Scatter plots of the components of the peculiar velocity $p$ (in units of 370 km s$^{-1}$) computed for 422 directions ($10^\circ\times10^\circ$ grid points) lying in sky at various polar angles with respect to the pole position (a) RA$=274^{\circ}$, Dec$=-29^{\circ}$ (or $l=3^{\circ}, b=-6^{\circ})$, derived for the Quaia redshift data (b) RA$=168^{\circ}$, Dec$=-7^{\circ}$ (or $l=264^{\circ}, b=48^{\circ}$) for the CMB dipole. Circles (o) with error bars in either case represent values for bin averages of the peculiar velocity components, obtained for various $20^{\circ}$ wide slices of the sky, while the dashed line shows the corresponding least square fit of $\cos \psi$ to the bin average values. 
For scatter plots, all points that corresponds to directions in a hemisphere $\Sigma_1$, centred about the Galactic Centre, are shown by the symbol '1', while all points in the hemisphere $\Sigma_2$, centred about the Galactic Anticentre, are shown by the symbol '2'. 
\label{F2}}
\end{figure*}

%--------------%---------------------------------------------------
\section{Results and discussion}\label{S3}
%---------------------------------------------------
\begin{figure*}
\includegraphics[width=\linewidth]{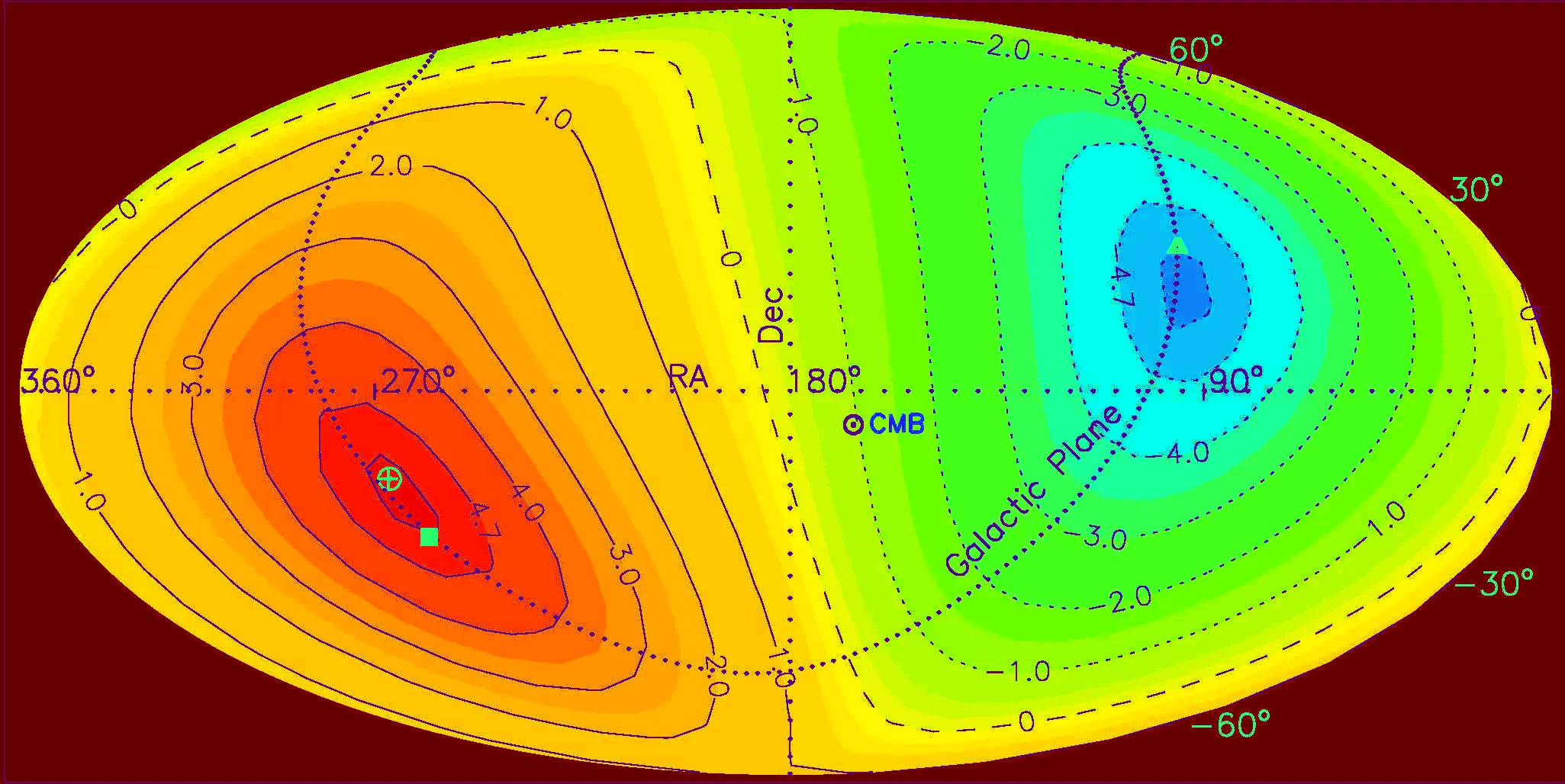}
\caption{A Hammer--Aitoff equal-area projection coloured map in equatorial coordinates, of the peculiar velocity components, estimated for 422 directions ($10^\circ\times10^\circ$ grid points) in the sky from the {\bf simulated} Quaia $m_{\rm G}<20.5$ redshift data. Overlaid are some representative contours, depicting amplitudes of the peculiar velocity $p$, in units of the CMB value of 370 km s$^{-1}$. The velocity vector assumed in this particular simulation was of amplitude $p=5$ in the direction of the Galactic Centre. The~horizontal and vertical axes denote RA and Dec. Positive component values are shown by continuous contour lines; the dashed curve represents zero amplitude of the peculiar velocity component while the negative component values are shown by dotted lines.  
The~optimum pole direction is expected to be in the vicinity of the highest contour value (in dark red colour) and is represented by the symbol $\oplus$ (in light blue). 
%lying at the position where a 3-d $\cos\psi$ fit yields a minimum chi-square.
Symbol $\odot$ indicates the CMB pole position. Also shown is the Galactic Plane (thick-dotted curve) with the symbol $\blacksquare$ (in light blue) indicating position of the Galactic Centre while $\blacktriangle$ (in light blue) indicates  position of the Galactic Anticentre. \label{F3}}
\end{figure*}
%---------------------------------------------------
\begin{figure*}
\includegraphics[width=\linewidth]{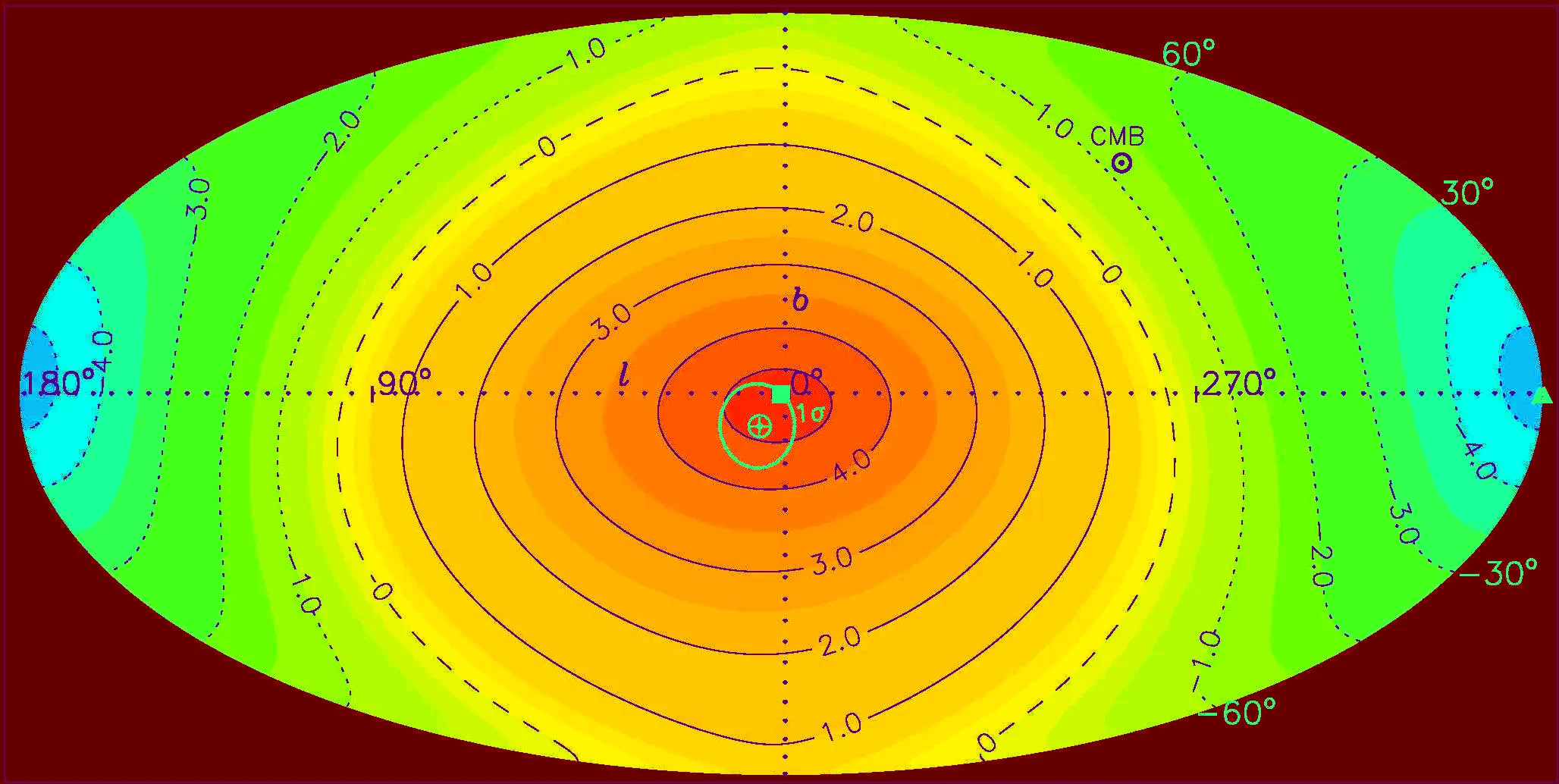}
\caption{A coloured map in Galactic coordinates (Hammer--Aitoff projection) of the components of Solar peculiar motion, estimated for 10360 directions ($2^\circ\times2^\circ$ grid points) in the sky from the  Quaia $m_{\rm G}<20$ redshift data. Overlaid are some representative contours, depicting amplitudes of the peculiar velocity, in units of the CMB value of 370 km s$^{-1}$. The~horizontal and vertical axes denote Galactic longitude ($l$) and latitude ($b$). 
Positive component values are shown by continuous contour lines; the dashed curve represents zero amplitude of the peculiar velocity component while the negative component values are shown by dotted lines.   
The~optimum pole direction, expected to be in the vicinity of the highest contour value (in dark red colour), is indicated by the symbol $\oplus$ (in light blue). 
%at a position where a 3-d $\cos\psi$ fit yields a minimum chi-square. 
It seems to lie in close proximity of the Galactic Centre indicated by $\blacksquare$ (in light blue) at $l = 0$, $b = 0$ that lies within $1\sigma$ errors, indicated by a  light blue-coloured circle around the symbol $\oplus$. The symbol $\odot$ indicates the CMB pole position. Symbol $\blacktriangle$ (in light blue) indicates  position of the Galactic Anticentre.\label{F4}}
\end{figure*}

Figure~\ref{F1} shows in equatorial coordinates a coloured Hammer--Aitoff equal-area projection map of the peculiar velocity components from the whole ($m_{\rm G}<20.5$) Quaia sample. 
Also overlaid are some representative contours, depicting amplitudes of the peculiar velocity, in units of the CMB value of 370 km s$^{-1}$. For the peculiar velocity vector, the maximum occurs at RA$=274^{\circ}$, Dec$=-29^{\circ}$, 
which seems to be pointing within $1\sigma$ errors, quite unexpectedly, around the direction of the Galactic Centre. Moreover, the amplitude of the peculiar velocity seems to be much larger, by at least a factor of 4.5, than that inferred from CMB dipole anisotropy. It is clear that the peculiar velocity inferred from the dipole anisotropy in the redshift distribution of distant quasars does not lie in the direction of the CMB dipole position (RA$=168^{\circ}$, Dec$=-7^{\circ}$), indicated by the symbol $\odot$ in Fig.~\ref{F1} and is instead pointing in the direction of the Galactic Centre, almost at a right angle (more precisely at $94^\circ$, as can be calculated from the right-angled spherical triangle using the law of cosines) in sky from the CMB dipole position. It should be noted that no reference to any information about our galaxy (Milky Way) has entered into the computations of the Solar peculiar velocity which has used only the published redshifts and sky coordinates of the quasars \cite{St24}. In fact, we did not even use the galactic coordinate system and performed all calculations first in the equatorial coordinate system and only later recomputed in the galactic coordinate system to verify that in both coordinate systems we are indeed getting consistent results.

We have also made a 3-d $\cos\psi$ fit for each of the $n=422$ positions for the remaining $n-1$ $p$ values, and computed the chi-square value for each of these $n$ fits. 
A reduced Chi-squared ($\chi^2_\nu$) values for the 3-d cos fits computed for all 422 directions across the sky, shows a clear minimum  (Fig.~(\ref{F1a})), from which we infer the optimum direction of the observer's peculiar velocity in galactic coordinates at $l=3^{\circ}$ and $b=-6^{\circ}$ which in equatorial coordinates is RA = 174$^{\circ}$, and Dec =-29$^{\circ}$, the position of the maximum of peculiar velocity in Fig.~(\ref{F1}) .
%--------------------------------------------

In Fig.~(\ref{F2}) we have plotted components of the peculiar velocity as scatter plots of the peculiar velocity $p$ computed for directions ($10^\circ\times10^\circ$ grid points) lying in sky at various polar angles with respect to the pole position (a) RA$=274^{\circ}$, Dec$=-29^{\circ}$ (or $l=3^{\circ}, b=-6^{\circ}$), derived for the Quaia redshift data (b) RA$=168^{\circ}$, Dec$=-7^{\circ}$ (or $l=264^{\circ}, b=48^{\circ}$) for the CMB dipole. For scatter plots in Fig.~(\ref{F2}), all points that corresponds to directions in a hemisphere $\Sigma_1$, centred about the Galactic Centre, are shown by the symbol '1', while all points in the hemisphere $\Sigma_2$, centred about the Galactic Anticentre, are shown by the symbol '2'. Curiously, almost all directions in the hemisphere $\Sigma_1$ show positive values for the peculiar velocity component while directions in the hemisphere $\Sigma_2$ show negative values for the peculiar velocity component, consistent with the peculiar motion vector lying along the Galactic Centre.
We also computed bin averages of the velocity  component in $20^{\circ}$ wide slices of the sky by divided the sky into bins of $20^\circ$ width in polar angle about the above pole positions. For each %$j^{th}$ 
bin we determined the mean peculiar velocity 
%$\bar{p}_{\rm j}$ 
and its rms due to the spread in velocity within the bin. 
 Circles (o) with error bars in either case represent values for these bin averages for various $20^{\circ}$ wide slices of the sky, while the dashed line shows the corresponding least square fit of $\cos \psi$ to these bin average values. 
%Bevington \& Robinson 2003).
A least square fit of $\cos \psi$ to the bin average values (Fig.~(\ref{F2})) shows that the computed peculiar velocity ($p$) values for various sky points at polar angles ($\psi$) do follow a systematic $\cos\psi$ dependence, which is consistent with the expectation that in case of a genuine dipole, one expects a  $\cos\psi$ dipole pattern with respect to the true dipole direction in sky, with a minimum of a chi-square fit in that direction. 
From Fig.~(\ref{F2}), it is apparent that $\cos \psi$ makes a very good fit to the data for the derived redshift pole position $\alpha=274^{\circ}$, $\delta=-29^{\circ}$, while a redshift pole assumed at the CMB position, $\alpha=168^{\circ}$, $\delta=-7^{\circ}$, makes a rather poor chi-square fit. 
 
As mentioned earlier, we have made a large number of Monte Carlo simulations to test our method. In each of these simulations we kept the original quasar positions, so as to keep the sky coverage unchanged, however, after a random redistribution of the observed redshifts among quasars, effects of an assumed peculiar velocity vector was applied to each of the redistributed quasar redshifts. On this mock catalogue we employed our procedure to recover the peculiar velocity vector and compared it with the assumed input values. Figure~\ref{F3} shows in equatorial coordinates a coloured Hammer--Aitoff equal-area projection map of the peculiar velocity components computed from one such {\bf simulated} Quaia $m_{\rm G}<20.5$ data. The velocity vector assumed in this particular simulation was of amplitude $p=5$ in the direction of the Galactic Centre so as to have a realistic value of the redshift dipole, similar to as found in Fig.~\ref{F1}. It is evident from Fig.~\ref{F3} that the recovered velocity vector matches the assumed velocity vector, both in amplitude (with $\Delta p\approx 0$) and direction (as it lies very close to the Galactic Centre with $\Delta \alpha\approx 5^\circ,\Delta \delta\approx 12^\circ$). In fact, though details may differ, Fig.~\ref{F3} otherwise has a quite close resemblance to Fig.~\ref{F1}. This not only validated our method and computer routines, it also allowed us to make an estimate of the uncertainties in direction and amplitude of the peculiar velocity vector from the rms values of the differences in amplitude and direction between the recovered and assumed velocity vectors in a thousand different simulations. We performed 500 of the simulations in equatorial coordinates of the quasars and another 500 in galactic coordinate system. 

\begin{figure*}
\includegraphics[width=\linewidth]{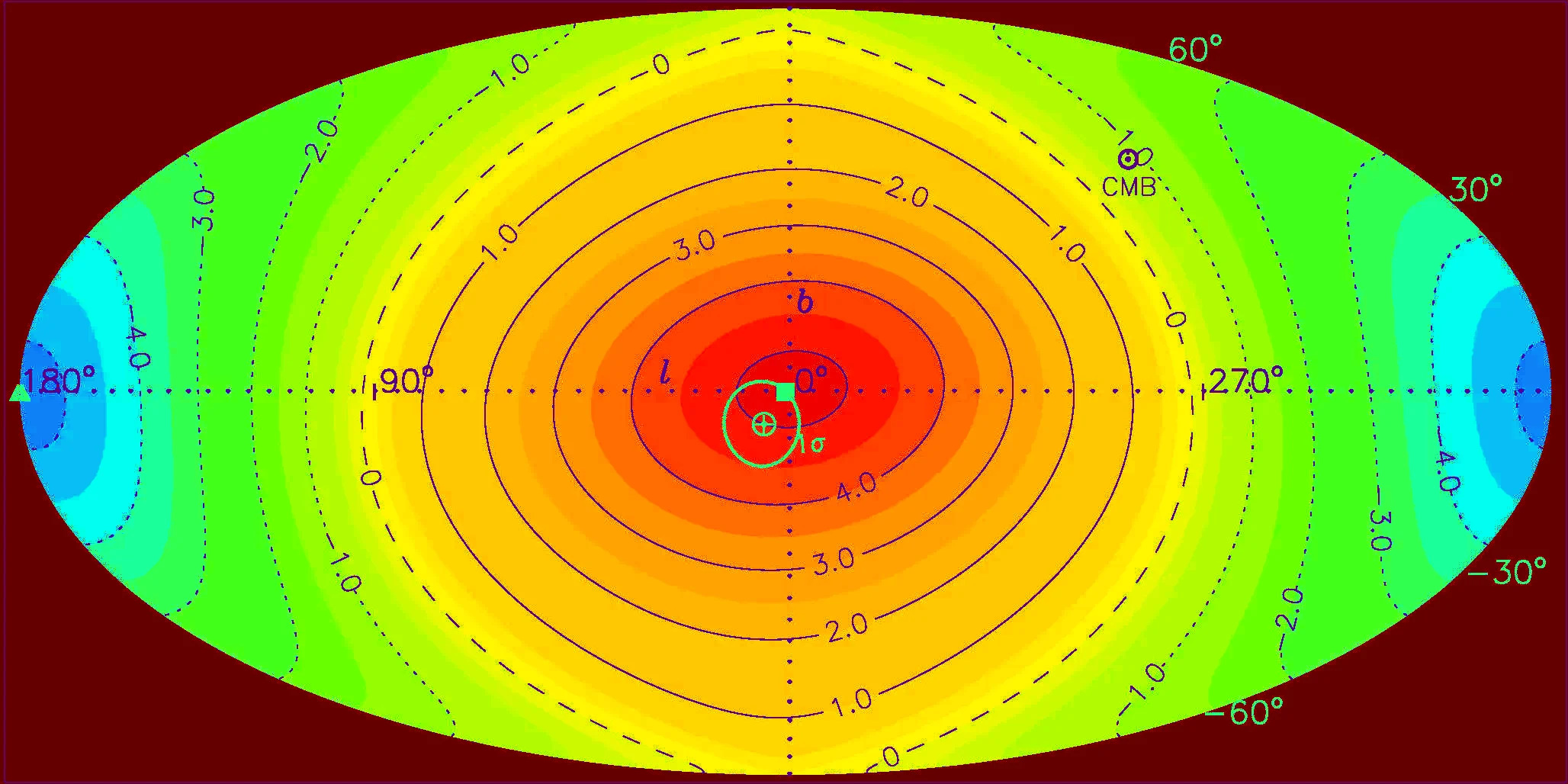}
\caption{A coloured map  in Galactic coordinates of the components of Solar peculiar motion, as in Fig.~\ref{F4} but from the Quaia $20<m_{\rm G}<20.5$ redshift data. 
%The centre displayed in Fig.~\ref{F5} has been shifted from Fig.~\ref{F4} by $90^\circ$, to exhibit clearly the layout of negative contours about the Galactic Anticentre. In the figure, $\Sigma_1$ represents the hemisphere centred about the Galactic Centre while $\Sigma_2$ represents the hemisphere centred about the Galactic Anticentre. 
\label{F5}}
\end{figure*}
%---------------------------------------------------
It has been emphasized that the brighter part of the Quaia sample ($m_{\rm G}<20$) may be cleaner than the full sample ($m_{\rm G}<20.5$) \cite{St24}. Therefore, we selected the brighter Quaia sample, with G-magnitudes $m_{\rm G}<20$, to determine the peculiar velocity to see whether we still get the same values. Figure~\ref{F4} shows a coloured map of the component of Solar peculiar motion, estimated for 10360 directions ($2^\circ\times2^\circ$ grid points) in Galactic coordinates, from the Quaia $m_{\rm G}<20$ redshift data.  
The~optimum pole direction, indicated by the symbol $\oplus$, where a 3-d $\cos\psi$ fit yields a minimum chi-square, seems to lie in close proximity of the Galactic Centre ($l = 0^\circ$, $b = 0^\circ$) that lies within $1\sigma$ errors. The inferred peculiar velocity is consistent with that obtained from the full sample $m_{\rm G}<20.5$ (Fig.~\ref{F1}).

We also employed the $20<m_{\rm G}<20.5$ sub-sample to estimate the peculiar velocity. Figure~\ref{F5} shows a coloured map of the component of Solar peculiar motion, estimated again for 10360 directions ($2^\circ\times2^\circ$ grid points) in Galactic coordinates, from the Quaia $20<m_{\rm G}<20.5$ redshift data. The redshift dipole in Fig.~\ref{F5} is very much consistent with that in  Fig.~\ref{F4}, though the data used in the two figures is with different magnitude limits without any overlap, with not even a single common source in the two sub-samples. Therefore, as far as the peculiar velocity determination from the redshift dipole is concerned, the fainter sub-sample ($20<m_{\rm G}<20.5$) may be as good as the brighter part of sample with $m_{\rm G}<20$.

%From Fig.~\ref{F5} it is apparent that the contours have a symmetry about the Galactic Centre. In order to indicate the hemisphere $\Sigma_1$, centred about the Galactic Centre ($l=0^\circ, b=0^\circ$), and the hemisphere $\Sigma_2$, centred about the Galactic Anticentre ($l=180^\circ, b=0^\circ$) unambiguously, we have displaced the centre in the displayed Fig.~\ref{F5} by $90^\circ$. This also allows us to see the layout of negative contours about the Galactic Anticentre more clearly. All positive contours seem to lie almost exclusively in the hemisphere $\Sigma_1$ while the negative contours are in the hemisphere $\Sigma_2$, with the zero value contour at the boundary between the two hemispheres. Such a symmetry though present in Fig.\ref{F1}, might not be immediately obvious as the Fig.\ref{F1} is drawn in equatorial coordinates. Actually, 
In all three figures (Figs.~\ref{F1}, \ref{F4} and \ref{F5}) it appears that in the Quaia data there is a systematic deficiency of quasar redshifts toward the direction of the Galactic Centre as compared to that toward the Galactic Anticentre direction, which may be due to the Solar peculiar motion in the Galactic Centre direction. 
%%-------------------------------- ------------
\begin{sidewaystable}
\begin{center}
\caption{\label{T1}Peculiar velocity estimates for the Quaia sub-samples for various G-magnitude ranges and different sky-coverage limits.} 
\hskip4pc\vbox{\columnwidth=33pc
\begin{tabular*}{\textwidth}{@{\extracolsep\fill}cccccccccc}
%\footnotesize
%\begin{tabular*}{ccccccccc}
\toprule%
%\hline 
(1)&(2)&(3)& (4)& (5)&(6)&(7)&(8)&(9)&(10)\\
Serial&G-magnitude & sky-coverage   & Quasar number &\multicolumn{4}{c}{Derived peculiar velocity direction} &\multicolumn{2}{c} {Peculiar velocity amplitude} \\
\cmidrule{5-8}\cmidrule{9-10}
% & ${\cal D}$& ${\cal D}_{\rm h}$  &  & $p_{\rm h}$\\
No.&($m_{\rm G}$) range & $|b|$ or $|\delta|$ limit  & in the sample &   $\alpha$ & $\delta$ & ${\it l}$ & ${\it b}$ &$p$ & $v$ \\% &  ($10^{-2}$)  \\% &  ($10^{3}$)

&&($^{\circ}$)&N&  ($^{\circ}$)&  ($^{\circ}$)&($^{\circ}$)&  ($^{\circ}$)&($370$ km s$^{-1}$) & ($10^{3}$ km s$^{-1}$)\\
\midrule
1&$m_{\rm G}< 20.5$   & $|b|>0$  &  1295502 &  $274\pm 08$ &  $-29\pm 08$ &  $03\pm 08$ &  $-06\pm 08$ & $4.6\pm0.8$ & $1.7\pm0.3$ \\

2&$m_{\rm G}< 20.5$   & $|b|>10$  &  1279488 &  $278\pm 08$ &  $-23\pm 08$ &  $10\pm 08$ &  $-06\pm 08$ & $4.9\pm0.8$ & $1.8\pm0.3$ \\

3&$m_{\rm G}< 20.5$   & $|b|>20$  &  1142085 &   $278\pm 09$ &  $-20\pm 09$ &  $13\pm 09$ &  $-05\pm 09$ & $3.8\pm0.8$ & $1.4\pm0.3$ \\

4&$m_{\rm G}< 20.5$   & $|b|>30$  & 917565 &  $280\pm 10$ &  $-23\pm 10$ &  $11\pm 10$ &  $-08\pm 10$ & $2.8\pm1.1$ & $1.05\pm0.4$ \\

5&$m_{\rm G}< 20.5$   & $|b|<30$  & 377937 &  $285\pm 12$ &  $-27\pm 12$ &  $09\pm 12$ &  $-14\pm 12$ & $11.0\pm6.0$ & $4.1\pm2.2$ \\\\

6&$m_{\rm G}< 20.0$  & $|b|>0$  &  755850 &   $277\pm 08$ &  $-29\pm 08$ &  $04\pm 08$ &  $-08\pm 08$ & $4.5\pm0.8$ & $1.65\pm0.3$ \\

7&$m_{\rm G}< 20.0$  & $|b|>10$   & 744833  &  $281\pm 08$ &  $-21\pm 08$ &  $13\pm 08$ &  $-08\pm 08$ & $4.7\pm0.8$ & $1.75\pm0.3$ \\

8&$m_{\rm G}< 20.0$  & $|b|>20$   & 660630  &  $284\pm 10$ &  $-17\pm 09$ &  $18\pm 10$ &  $-09\pm 10$ & $3.2\pm0.9$ & $1.2\pm0.35$ \\

9&$m_{\rm G}< 20.0$  & $|b|>30$   & 530364  &  $281\pm 10$ &  $-19\pm 10$ &  $15\pm 10$ &  $-07\pm 10$ & $2.0\pm1.4$ & $0.75\pm0.5$ \\

10&$m_{\rm G}< 20.0$  & $|b|<30$   & 225486  &  $292\pm 14$ &  $-29\pm 14$ &  $10\pm 14$ &  $-20\pm 14$ & $11.5\pm6.5$ & $4.3\pm2.4$ \\\\

11&$20<m_{\rm G}<20.5$  & $|b|>0$  &  539688 &  $270\pm 09$ &  $-29\pm 08$ &  $02\pm 09$ &  $-03\pm 08$ & $4.9\pm0.9$ & $1.8\pm0.35$ \\

12&$20<m_{\rm G}<20.5$  & $|b|>10$  & 534689 &  $273\pm 09$ &  $-25\pm 09$ & $06\pm09$ & $-03\pm09$ & $5.2\pm1.0$ & $1.9\pm0.4$ \\

13&$20<m_{\rm G}<20.5$  & $|b|>20$  & 481483 &  $272\pm 10$ &  $-24\pm 10$ & $07\pm10$ & $-02\pm10$ & $5.5\pm1.1$ & $2.0\pm0.4$ \\

14&$20<m_{\rm G}<20.5$  & $|b|>30$  & 387222 &  $276\pm 10$ &  $-23\pm 10$ & $09\pm10$ & $-05\pm10$ & $4.6\pm1.2$ & $1.7\pm0.45$ \\

15&$20<m_{\rm G}<20.5$  & $|b|<30$  & 152466 &  $270\pm 14$ &  $-21\pm 14$ & $08\pm14$ & $+01\pm14$ & $10.0\pm5.4$ & $3.7\pm2.0$ \\\\

16&$m_{\rm G}< 19.5$   & $|b|>0$  &  396024 &   $282\pm 14$ &  $-25\pm 10$ &  $10\pm 13$ &  $-10\pm 11$ & $4.6\pm1.3$ & $1.7\pm0.5$ \\

17&$m_{\rm G}< 19.0$  & $|b|>0$  &  185819 &   $298\pm 14$ &  $-29\pm 19$ &  $12\pm 15$ &  $-25\pm 19$ & $6.5\pm1.5$ & $2.4\pm0.55$ \\

18&$m_{\rm G}< 18.5$   & $|b|>0$  &  78450 &   $301\pm 34$ &  $-21\pm 23$ &  $21\pm 27$ &  $-25\pm 26$ & $7.3\pm2.6$ & $2.7\pm0.95$ \\

19&$m_{\rm G}< 20.5$   & $|\delta|>30$  &  619581 &   $241\pm 12$ &  $-35\pm 12$ &  $342\pm 12$ &  $+13\pm 12$ & $4.2\pm0.9$ & $1.55\pm0.35$ \\

20&$m_{\rm G}< 20.5$   & $|\delta|<30$  &  675921 &   $289\pm 12$ &  $-13\pm 12$ &  $24\pm 12$ &  $-11\pm 12$ & $4.4\pm0.9$ & $1.6\pm0.35$ \\
\botrule
%----------------------------------------
%\hline
\end{tabular*}
%\normalsize
%\end{tabular*}
}
\end{center}
\begin{flushleft}
\footnotetext{Various columns in Table~\ref{T1} are arranged as follows: Column (1) gives the serial number of each entry. Column (2) lists the G-magnitude ($m_{\rm G}$) range of the sub-sample. Column (3) gives the galactic latitude ($|b|$) or declination ($|\delta|$) limits. Column (4) lists the number of quasars in the sub-sample. Columns (5) and (6) list the direction of the peculiar solar velocity in equatorial coordinates ($\alpha$ and $\delta$). Columns (7) and (8) give the direction in  galactic coordinates ($l$ and $b$). Column (9) lists  $p$, the peculiar speed, in units of the CMB value 370 km s$^{-1}$, estimated for the corresponding sub-sample. Column (10) gives the peculiar solar velocity $v$, in units of $10^{3} $ km s$^{-1}$.}
\end{flushleft}
\end{sidewaystable}
%%-------------------------------- ------------

Though it would be desirable to have our quasar sample covering as many different directions as possible in sky, however, it is not absolutely essential to have a uniform distribution over the sky, as the method does not compare the number density of quasars in various directions. All we require is that there should be no direction-dependent bias introduced in the redshift determinations and that a considerable  number of quasars with known redshifts in various directions are present in the sample so that we can achieve a reasonable statistical estimate of the contribution of our peculiar motion to the observed redshift as a function of polar angle with respect to the direction of peculiar velocity vector (Eq.~\ref{eq:2}). Therefore we can apply large cuts in galactic latitude or even in declination and still determine the peculiar velocity. Of course with reduction in the sample size, statistical uncertainties might increase, but hopefully without any systematic effects. Results for the peculiar velocity, with various cuts in the sky coverage are presented in Tables~\ref{T1}.

In Table~\ref{T1} we have summarized not only the results that have been presented earlier and shown in Figs.~\ref{F1}, \ref{F4} and \ref{F5}, we also present the peculiar velocity values determined from data with various cuts in galactic latitude or in declination. We have also employed some sub-samples with additional limits on G-magnitude, of course all belonging to the Quaia sample, and presented the results in rows with serial Numbers 16-18 in Table~\ref{T1}. Consistency of the results with various cuts in the sky coverage or in the G-magnitudes is quite apparent; particular attention may be paid to comparison of sub-samples in rows with serial Numbers 6-10 ($m_{\rm G}<20$) with sub-samples in serial Numbers 11-15  ($20<m_{\rm G}<20.5$) having no overlaps and thus no common sources, but they all within errors are giving consistent values for the peculiar velocity, both in direction and amplitude. Similarly is the case with the sub-samples in serial Numbers 4 ($|b|>30^\circ$) and 5 ($|b|<30^\circ$), or 9 and 10, or 14 and 15, which also have no  overlaps and thus no common sources. We can also compare the  sub-samples in rows with serial Numbers 19 ($|\delta|>30^\circ$) and 20 ($|\delta|<30^\circ$), which again have no overlap of sources and still yield similar peculiar velocity values.

In a recent analysis of a colour dependence of dipole in CatWISE2020 data, it was seen that while
the dipole in one colour bin points close to the direction of the CMB dipole, the dipole in
the other colour bin, which apparently is at a relatively higher redshift, points in a direction quite close to the Galactic Plane, although in a direction roughly opposite to the Galactic Centre \cite{Pa24}. However, it is quite apparent that the number density dipole does not coincide with the redshift dipole as even in the Quaia sample, where the dipole estimated from dipole asymmetries in number counts \cite{Si24b} points approximately toward the CMB dipole, the redshift dipole points toward the Galactic Centre, an almost $90^\circ$ away direction in sky. We have performed sufficient checks to ensure that our procedure and routines are not causing this unexpected and unusual result and that this redshift dipole is genuinely present in the Quaia data.

One point that could be of concern here is whether some galactic position-dependent systematics have crept into the Quaia quasar redshifts. This could be an important issue, more so because the Quaia quasar sample is constructed from the Gaia DR3 quasar candidates sample \cite{Ga23b}, found by the Gaia satellite while performing its all-sky survey of our Galaxy. Gaia survey mission basically was designed and optimized to obtain astrometry, photometry, and spectroscopy of nearly two billion stars in the Milky Way \cite{Ga16}. Some systematics could enter, e.g., because of attenuation due to dust, near the Galactic plane, the relative number density of quasars  observed at different galactic coordinates might get affected. For one thing, and as has been mentioned earlier, a mere differential number density of quasars in a certain direction, cannot cause a dipole in redshift distribution in that direction. In fact, quasars in the Quaia sample seem to possess a number-density dipole, 3-4 times larger than the CMB dipole, in the same direction as the CMB dipole \cite{Si24b}, and which, as we mentioned earlier, is in a direction at almost a right angle to the Galactic Centre.  

%(Gaia Collaboration et al. 2016) %also observed millions of extragalactic objects.  
The Quaia sample, in fact, has been asserted to be a highly competitive sample for cosmological large-scale structure analyses \cite{St24}. Random errors in redshift measurements would not give rise to redshift contours to possess the observed $\cos\theta$ dependence about the Galactic Centre, that too in all sub-samples (Table~\ref{T1}), as the signature of the Solar peculiar velocity pointing along the Galactic Centre is seen in various G-magnitude bands as well as at different galactic latitudes. 
Now, if a redshift dipole in the direction of the Galactic Centre indeed appears because of systematics  present in the Quaia catalogue,  this would, in turn, indicate that the original Gaia data contains such systematics since the Quaia data are derived from those. Only if a bias is present in quasar redshift distribution with respect to the Galactic plane, for example, if the attenuation due to dust near the galactic plane were to affect the redshift distribution of sources selected at different galactic latitudes differentially, it might possibly give rise to a spurious redshift dipole pointing towards the Galactic Centre, 
However, such systematics are unlikely to give rise to {\it similar} redshift dipoles from data with cuts applied at high as well as low galactic latitudes (compare e.g., sub-samples with $|b|>30^\circ$ and $|b|<30^\circ$ in Table 1 with serial Numbers 4 and 5, or 9 and 10, or 14 and 15). 

At the same time, from the existing Quaia data, one could not completely reject the possible presence of a genuine redshift dipole pointing towards the Galactic Centre, even if it seems to go against the conventional wisdom, after all data act as the supreme judge, though one should still be wary of a bias inadvertently creeping somehow into the data. The only way to confirm or rule out such a scenario would perhaps be a future comparison with an equivalent or even better, quasar catalogue with known sky positions and independently derived redshifts. 
%with   
% (Section~\ref{S3}),
% However, there is no hint of such a systematic bias entering into the redshift estimation of the Quaia sample when being formed from the original Gaia DR3 quasar candidate sample \cite{Ga23b} and is,  
%Therefore there is no reason to suspect that the unusual Solar peculiar motion derived here from the redshift dipole in the Quaia sample is a result of some such systematics. 
%and another independent large quasar sample, with measured redshifts, may be required to further confirm the pointing direction of the redshift dipole.

In the conventional picture the Solar system has a rotational motion around the Milky Way of a value 220-250 km s$^{-1}$ \cite{Sh82,Ca07}. A recent estimate of the Solar system motion gave 250 km s$^{-1}$ as the azimuthal component around the Galactic Centre and 14 km s$^{-1}$ as the radial component of motion toward the Galactic Centre \cite{Sc12}. These estimates use stellar kinematics in the Milky Way and depend upon various assumptions about the shape and structure of the Milky Way.
On the other hand, we derived the Solar system peculiar velocity, $\sim 1700$ km s$^{-1}$ in a direction toward the Galactic Centre, from the dipole asymmetry in the redshift distribution of 1.3 million distant quasars, without referring to any information about the galactic structure. 
The only assumption in our method is that of CP, according to which distant quasars should have an  isotropic distribution of cosmological redshifts as seen by an observer stationary in comoving coordinates of the expanding universe and that the dipole anisotropy, if any, observed in their redshift distribution is due to a peculiar motion of the observer along with the Solar system. 
Of course, we cannot ascertain whether the motion of the Solar system with a large speed of $\sim 1700$ km s$^{-1}$ is within the Milky Way toward its centre or is it that the Solar system along with the whole Milky Way galaxy, perhaps part of a larger unit comprising clusters of galaxies, is moving in a direction which, as seen from a vantage point on Earth, happens to coincide with the direction of the Milky Way centre. In either case, the peculiar motion derived from the observed dipole anisotropy in the redshift distribution of 1.3 million distant quasars in sky does not seem to agree with that derived conventionally from a CMB dipole asymmetry. Though such large value of the Solar peculiar speed have been inferred from the dipole anisotropies in the sky brightness arising from discrete sources or/and their number counts in the sky, approximately along the same direction as the CMB dipole \cite{4,5,6,7,8,9,11,Si21,Sie21,Se22,Wa23,Si24a,Si24b,Pa24}, 
%(Singal 2011; Rubart \& Schwarz 2013; Tiwari et al. 2015; Colin et al. 2017; Bengaly et al. 2018; Singal 2019a,b; Secrest et al. 2021; Siewert et al. 2021; Singal 2021a,b) 
a direction that is almost at a right angle to the  redshift dipole direction derived here.

Significantly larger values for peculiar velocity, inferred from anisotropies observed in number counts of active galactic nuclei (AGNs) surveys \cite{4,5,6,7,8,9,11,Si21,Sie21,Se22,Wa23,Si24a,Si24b,Pa24} as compared to the CMB dipole, though almost always with direction along the CMB dipole, do not support a peculiar motion of the Solar system to be their common cause, including that for the CMB dipole. After all the Solar system peculiar velocity cannot depend upon the particular data or methodology used. Now, various AGN dipoles pointing toward approximately the same direction in the sky cannot be due to some random fluctuations or some ill-understood systematics in the data or the techniques, otherwise the dipoles should be pointing in directions randomly distributed in sky as many of these have been derived by many independent groups employing their own routines from different datasets. 

What one actually observes in these catalogues is the presence of a dipole anisotropy in the number density of sources \cite{4,5,6,7,8,9,11,Si21,Sie21,Se22,Wa23,Si24a,Si24b,Pa24}, toward the CMB dipole direction (although with an amplitude much larger than the CMB dipole) and which might be interpreted, in accordance with the CP, resulting from observer's peculiar motion as part of the Solar system \cite{15}. But then one should have also seen this component of the Solar motion toward the CMB dipole direction in the Quaia quasar redshifts, irrespective of the true nature of the observed redshift dipole toward the Galactic Centre, which is not seen since the observed redshift dipole is  at $\sim 90^\circ$ from the CMB dipole direction. An absence of such a dipole component in the quasar redshifts toward the CMB dipole direction may indeed imply that Solar motion is not a cause of these dipoles and which may further be a pointer to an inherent presence of dipole anisotropies in number counts, in contravention of the CP.

%And for that to happen, the present results from Quaia sample, which appear against the conventional wisdom, also must be brought to the attention of the concerned scientific community.

%In such a scenario, the dipole derived from quasar redshifts may have no relation with the dipoles derived from the AGN number counts. 
%(Singal 2011; Rubart \& Schwarz 2013; Tiwari et al. 2015; Colin et al. 2017; Bengaly et al. 2018; Singal 2019a,b; Secrest et al. 2021; Siewert et al. 2021; Singal 2021a,b) 
Now, if the CP itself is in doubt, then even the isotropic distribution of cosmological redshifts of quasars cannot be assumed and the redshift dipole may also be not due to observer's peculiar motion implying the presence of genuine anisotropy, i.e. an intrinsic dipole, in the quasar redshifts, 
However, in this case we are not only denying the CP, a cornerstone in the basic foundation of the standard model in cosmology,  
there is an equally enigmatic question that why among quasars, presumed to be at cosmological distances as inferred from their redshifts, there exists such a dipole in their redshift distributions that coincides with the direction of the centre of a relatively tiny system, like the Milky Way, as seen   from the vantage point of an observer in the Solar system. 
%\section{DISCUSSION}\label{S5}
\section{Conclusions}\label{S4}
A redshift anisotropy in a large all-sky sample of 1.3 million quasars, has yielded a Solar system peculiar velocity which is 4-5 times larger than the peculiar motion inferred from a dipole asymmetry in the CMB. Even more curiously, direction of the inferred peculiar velocity within $1\sigma$ coincides with that of the Galactic Centre. Of course, it cannot be ascertained whether we are moving with a large speed of $\sim 1700$ km s$^{-1}$ toward the Milky Way centre or it is the whole  Milky Way galaxy, perhaps part of a larger system, is moving in a direction which, as seen from a vantage point on Earth, happens to coincide with the direction of the Milky Way centre. In either case, the peculiar motion derived from the  observed asymmetry in the redshift distribution of 1.3 million quasars in sky does not seem to agree with that derived conventionally from a CMB dipole asymmetry; a result which is not in accordance with the cosmological principle, a cornerstone in the basic foundation of the standard model in cosmology.
%Both alternatives are fraught with
%%%%%%%%%%%%%%%%%%%%%%%%%%%%%%%%%%%%%%%%%%
%\newpage
\section*{Declarations}
The author has no conflicts of interest/competing interests to declare that are relevant to the content of this article. 
\section*{Funding}
No funds or grants were received from anywhere for this research.
%%%%%%%%%%%%%%%%%%%%%%%%%%%%%%%%%%%%%%%%%%
%--------------------------------------------
\section*{Data Availability}
The data used in this article is freely available at DOI 10.5281/zenodo.8060755. 
%--------------------------------------------

\end{document}